\documentclass[sigconf]{acmart}

\AtBeginDocument{%
  \providecommand\BibTeX{{%
    \normalfont B\kern-0.5em{\scshape i\kern-0.25em b}\kern-0.8em\TeX}}}

\setcopyright{acmcopyright}
\copyrightyear{2023}
\acmYear{2023}
\setcopyright{rightsretained}
\acmConference[CHI '23]{Proceedings of the 2023 CHI Conference on Human Factors in Computing Systems}{April 23--28, 2023}{Hamburg, Germany}
\acmBooktitle{Proceedings of the 2023 CHI Conference on Human Factors in Computing Systems (CHI '23), April 23--28, 2023, Hamburg, Germany}\acmDOI{10.1145/3544548.3581357}
\acmISBN{978-1-4503-9421-5/23/04}

\graphicspath{{Graphics/}}

\usepackage{multirow}
\usepackage{booktabs}
\usepackage{caption}
\usepackage{subcaption}
\usepackage{enumitem}
\usepackage{float}
\usepackage{tabularx}
\newcolumntype{Y}{>{\centering\arraybackslash}X}
\usepackage{color, xcolor,colortbl}
\definecolor{shade}{rgb}{0.9,0.9,0.9}

\begin{document}

\title[Cross-Race Psychological Responses to Failures of Automatic Speech Recognition]{Can Voice Assistants Be Microaggressors?\\ Cross-Race Psychological Responses to \\Failures of Automatic Speech Recognition}

\author{Kimi V. Wenzel}
\affiliation{{%
  \institution{Carnegie Mellon University}
  \streetaddress{5000 Forbes Avenue}
  \city{Pittsburgh}
  \state{Pennsylvania}
  \country{USA}
  \postcode{15213}
}}
\email{kwenzel@cs.cmu.edu}

\author{Nitya Devireddy}
\affiliation{{%
  \institution{Carnegie Mellon University}
  \streetaddress{5000 Forbes Avenue}
  \city{Pittsburgh}
  \state{Pennsylvania}
  \country{USA}
  \postcode{15213}
}}
\email{ndevired@alumni.cmu.edu}

\author{Cam Davidson}
\affiliation{{%
  \institution{Carnegie Mellon University}
  \streetaddress{5000 Forbes Avenue}
  \city{Pittsburgh}
  \state{Pennsylvania}
  \country{USA}
  \postcode{15213}
}}
\email{jcdaviso@alumni.cmu.edu}

\author{Geoff Kaufman}
\affiliation{{%
  \institution{Carnegie Mellon University}
  \streetaddress{5000 Forbes Avenue}
  \city{Pittsburgh}
  \state{Pennsylvania}
  \country{USA}
  \postcode{15213}
}}
\email{gfk@cs.cmu.edu}

\renewcommand{\shortauthors}{Wenzel et al.}


\begin{abstract}
Language technologies have a racial bias, committing greater errors for Black users than for white users. However, little work has evaluated what effect these disparate error rates have on users themselves. The present study aims to understand if speech recognition errors in human-computer interactions may mirror the same effects as misunderstandings in interpersonal cross-race communication. In a controlled experiment (N=108), we randomly assigned Black and white participants to interact with a voice assistant pre-programmed to exhibit a high versus low error rate. Results revealed that Black participants in the high error rate condition, compared to Black participants in the low error rate condition, exhibited significantly higher levels of self-consciousness, lower levels of self-esteem and positive affect, and less favorable ratings of the technology. White participants did not exhibit this disparate pattern. We discuss design implications and the diverse research directions to which this initial study aims to contribute.
\end{abstract}

\ccsdesc[500]{Human-centered computing~Empirical studies in HCI}
\ccsdesc[500]{Social and professional topics~Race and ethnicity}

\keywords{Language Technology; Voice Assistants; Conversational User Interface; Automated Speech Recognition; Wizard-of-Oz; Race; Microaggressions; Harm; Individual Differences; Quantitative Methods}

\maketitle

\section{Introduction}
Language technologies are growing in both presence and power. By 2024, voice assistants (VAs) like Apple Siri, Google Assistant, and Amazon Alexa are expected to be accessible on over 8.4 billion devices worldwide  \cite{juniper}. While  these technologies are becoming increasingly ubiquitous in everyday life, assisting in tasks from the mundane (e.g., asking about current weather conditions) to the significant (e.g., calling for help in an emergency), they do not yet serve all users equally well. One population that is particularly poorly served by the speech recognition technology that powers VAs is Black American users. A growing body of work has demonstrated that word error rates in automated speech recognition systems are significantly higher for Black users than for white users,\footnote{We capitalize Black but not white, per the reasoning set by Kong \cite{kong2022intersectionally}.} a pattern largely attributed to the fact that Black voices are underrepresented in the voice samples that comprise the datasets on which these technologies are programmed \cite{koenecke2020racial, tatman2017effects}. In this paper, we argue that such errors, beyond merely limiting the function and utility of VAs for Black users, may also be experienced as \textit{microaggressions}, subtle acts of bias that reinforce marginalization or the feeling of being ``othered'' in social interactions. Building on prior work demonstrating that misunderstandings in cross-race interactions are often coded by racial minority groups as microaggressions, as well as past work demonstrating that people treat computers as social entities, we predicted that Black users would exhibit similar patterns of responses to speech recognition errors exhibited by VAs. Specifically, we tested what psychological harm might be caused by these errors in the immediate aftermath of an encounter with an error-prone virtual assistant. 

While previous literature has demonstrated the detrimental effects microaggressions have on racial minorities \cite{williams2020psychology}, especially as it pertains to their mental health \cite{paradies2015racism}, little work has examined what impact a high word error rate specifically may have on individuals \cite{field2021survey, mengesha2021don}. This is reflective of broader research trends on bias in computer systems taking an act-based approach. Act-based approaches focus on identifying and measuring the forms in which systems discriminate (i.e. What are the acts of bias?). In contrast, \textit{harm-based} approaches measure the distinct effects and ways in which these biases harm impacted individuals (i.e. What are the harms of bias?) \cite{freeman2021toward, lippert2013born}. This lack of prior harm-based work is more than a knowledge gap; it perpetuates the continued decentering of people of color and their experiences, and a continued de-emphasis on the impact of inequitable and/or non-inclusive technologies. Thus, rather than focusing on act-based accounts of bias in speech recognition systems, as most previous work has done, we instead take a harm-based approach and study the psychological effects encountering speech recognition errors from a VA may have on Black users. 

We report the methods and findings from a controlled experiment, in which Black and white users were randomly assigned to interact with a VA designed to commit a high versus low rate of errors on a set of pre-designated tasks. We employed a set of psychometric outcome measures utilized in prior empirical research on microaggressions -- including measures of emotional response, self-consciousness, individual and group-level self-esteem, and overall evaluations of the VA -- to evaluate the psychological impact of word error rate on Black and white users. 

This paper makes the following novel research contributions:
\begin{itemize}
	\item We introduce a harm-based, microaggressions-centered framework to understand marginalized group members' interactions with   language technologies.
	\item We conducted the first controlled experiment with quantitative outcome measures of the impact of voice assistant errors as a type of microaggression toward Black users.
	\item We provide evidence that, compared to white users, voice assistant errors significantly \emph{raise} Black users' levels of self-consciousness.
	\item We provide evidence that, compared to white users, voice assistant errors significantly \emph{lower} Black users' mood, individual self-esteem, collective self-esteem, and their evaluation of voice assistant technologies.
	\item We outline several approaches to designing for harm mitigation and coping with technology-mediated microaggressions.
\end{itemize}

\section{Related Work}

\subsection{Bias and Accuracy Degradation in Language Technology}
Previous work has demonstrated that the accuracy of language technology degrades for certain demographic groups. For example, in one evaluation, Twitter's language identifier marked tweets using African American English as a foreign language 19.7\% more than tweets using white-aligned English \cite{blodgett2017racial}. And in online hate speech detection, a false positive bias has been consistently demonstrated toward African American English \cite{sap2019risk}. 

Regarding automated speech recognition, VA users with foreign accents are more likely to experience errors \cite{palanica2019you}. Such errors even extend to natives without foreign accents: In a study of YouTube's automated captions, Tatman \textit{et al.} found that captions for Black speakers were significantly less accurate than that of their white counterparts \cite{tatman2017effects}. Most notably, Koenecke \textit{et al.} found speech recognition systems of Amazon, Apple, Google, IBM, and Microsoft to have an average word error rate of 35\% for Black American speakers, in contrast to a 19\% word error rate for white American speakers \cite{koenecke2020racial}. While such accuracy degradation has been repeatedly established, little work has taken a harm-based approach and been devoted to understanding precisely what effect accuracy degradation in VA systems has on users themselves. Mengesha \textit{et al.} conducted a diary study evaluating Black users' subjective experiences with voice assistants, including their responses to VA errors. This study revealed powerful testimonials about Black users' perceptions of language technologies, including the perception that such technologies are not designed with Black users in mind and require some degree of speech accommodation in order to function well for Black users \cite{mengesha2021don}. The present study builds on this work by using controlled experimental methods to more precisely measure the the psychological effects of experiencing those shortcomings in the technology. To our knowledge, the present study is the first systematic evaluation of the psychological impact of automated speech recognition errors that utilizes an experimental design and quantitative measurement methods.

\subsection{The Experience and Impact of Racial Microaggressions and Stereotype Threat}

Racial microaggressions are defined as ``brief and commonplace daily verbal, behavioral, or environmental indignities, whether intentional or unintentional, that communicate hostile, derogatory, or negative racial slights and insults toward people of color'' \cite{sue2007racial}. According to psychologist Derald Wing Sue, microaggressions represent a primary form of ``modern racism,'' subtle and often invisible forms of prejudice or inequity ``hiding in the invisible assumptions and beliefs of individuals'' and  ``embedded in the policies and structures of our institutions'' \cite{sue2010microaggressions}. Microaggressions commonly arise in conversational contexts, in which intergroup differences can manifest in the verbal or nonverbal responses of interaction partners from more privileged identity groups. Indeed, people of color identify the common experience of being ignored, being asked to repeat themselves, and/or encountering misunderstandings from white conversation partners due to differences in speech patterns or word choice and, specifically, any deviation from white American English\footnote{The authors have chosen the term ``white American English'' over the conventional ``Standard American English'' (SAE): Despite linguists' agreement that other language varieties, including African American English, are of equal legitimacy to white American English \cite{LSA}, the term ``SAE'' continues to be used in scholarly work \cite{coupland2000sociolinguistic}, to refer ``not coincidentally [to] the language of primarily white, middle- and upper middle-class, and middle-American communities'' \cite{lippi1997we}. While we acknowledge that ``white American English'' is an imperfect label, we aim to provoke the NLP community to reflect on raciolinguistic ideologies.} exhibited by people of color \cite{gomez2011microaggressions, huber2011discourses, mcclure2020escalating, minikel2013racism, sue2008racial}. 

Although microaggressions tend to be subtle in nature, and often unrecognized by those who commit them, they can have a profound effect on those who experience them. A simple, seemingly innocuous example of being misunderstood or being asked to repeat oneself because of the way one speaks can reinforce the salience of a marginalized identity. This is particularly likely when there are societal stereotypes that associate one's identity group with lower levels of intelligence and/or poorer communication skills \cite{ayala2020outing}. Specifically, microaggressions can trigger stereotype threat, a “socially premised psychological threat that arises when one is in a situation or doing something for which a negative stereotype about one's group applies” \cite{steele1995stereotype}. Prior work has shown that stereotype threat can have a host of psychological effects, including increased cognitive load \cite{croizet2004stereotype} and self-focus \cite{brown2003stigma}, increased anxiety \cite{osborne2007linking}, and decreased self-esteem \cite{crocker1989social}. Moreover, stereotype threat can hinder targets' subjective experiences \cite{adams2006detrimental}, lower their sense of belonging \cite{walton2007question}, and cause them to dis-identify with or disengage from particular domains associated with the threat \cite{smith2007stereotyped}.

\subsection{Voice Assistants as Social Actors} 
Given that the effects of microaggressions among people in human-human interactions, specifically in occurrences of miscommunication and misunderstanding, are well-documented, the present work aimed to study if these effects may be mirrored in human-computer interactions. Nass \textit{et al.}'s \emph{Computers are Social Actors} paradigm affirms that people subconsciously apply social heuristics to  technologies, despite their conscious awareness that these technologies are not sentient \cite{nass1994computers}. This paradigm has been exhibited across multiple contexts: People form first impressions of a voice's ``personality'' \cite{mcaleer2014you} much like how they form first impressions of people \cite{albright1988consensus}, and are attracted to computer voices that demonstrate similar personality characteristics as themselves \cite{nass2001does, lee2000can} just as people are attracted to those who are similar to them \cite{montoya2013meta}. People also apply social codes of politeness towards voice assistants \cite{bonfert2018if}, much like how we employ politeness among other people \cite{brown1987politeness}. What's more, stereotyping gender-based attributes is commonplace for voice technology users: Computer tutors with characteristically male voices were rated more competent than female-voiced tutors \cite{nass1997machines}, in line with people's general perceptions of gender and competence \cite{eagly1982inferred, wood1986sex, eagly2002role}. In more recent work, researchers have found that some voice assistant users even actively personify modern assistants like Amazon Alexa and Google Home \cite{purington2017alexa, choi2021ok}. In short, ``humans have become \emph{voice-activated} with brains that are wired to equate voices with people and to act quickly on that identification'' regardless of whether the voice is artificial or representative of a real person \cite{nass2005wired}. 

\subsection{Hypotheses}
Building off of these established phenomena, we predict that the effects of microaggressions and stereotype threat demonstrated in interpersonal interactions will carry over to Black users' experience interacting with an error-prone VA. Specifically, we designed a controlled experiment to test the following hypothesis:

\begin{itemize}[font=\bfseries,
  align=left,topsep=3pt]
\item[H1:] Black users will exhibit a pattern of responses to speech recognition errors committed by a virtual assistant similar to the pattern previously demonstrated in research on racial microaggressions: \textbf{(a)} heightened self-consciousness; \textbf{(b)} lower levels of positive affect; \textbf{(c)} higher levels of negative affect (in particular, anxiety); \textbf{(d)} reduced individual self-esteem; \textbf{(e)} reduced collective self-esteem; and \textbf{(f)} more negative evaluations of the voice assistant.
\item[H2:] White users, in contrast, will not experience speech recognition errors as microaggressions and, thus, not be expected to exhibit this pattern of response
\end{itemize}

\section{Methodology}
All materials and procedures described below were approved by the institutional review board at the authors' university.

\subsection{Recruitment (N=108)}
A total of 108 participants were recruited through a call for study participants on the following Craigslist city pages: Atlanta (n=21), Chicago (n=21), Houston (n=22), New York (n=22), and Washington D.C. (n=22). This sample size was determined using a power analysis based on a predicted effect size of .69, as informed by prior meta-analyses of research documenting the psychological harm of microaggressions. Participants were screened for eligibility before beginning the experiment. Requirements for eligibility included residing in the U.S.A., being aged 18 or older, identifying as either Black or white, having access to a device with a microphone and web camera, and being an active user of voice technology (using a voice technology ``multiple times a day'' or ``multiple times a week''). We required participants to be active users of voice technology to minimize friction in the beginning of the study procedure, which was especially important given that the study was conducted over Zoom. Furthermore, this requirement helped streamline the procedure such that participants had as little direct interaction with the researchers as possible.

The results that follow are based off of the responses of 108 participants, 54 who identified as Black and 54 who identified as white. To determine participants' race, in the screener form participants were asked to select from a set of race and ethnicity items in response to the question: ``What is your race/ethnicity? Please select all that apply.'' Only participants who indicated they were ``African American/Black'' or ``White/Caucasian'' were invited to participate. Mixed race individuals were not included. The mean age of the participants was 25.7, with an age range 18-52. 48 participants identified as male, 48 identified as female, and 12 identified as another gender or did not specify. All participants were compensated USD 15. 

\subsection{Study Design and Procedure}
The study utilized a 2x2 between-subjects design, with participants' self-identified race (Black, white) and their randomly assigned error rate condition (low, high) representing the two independent variables of interest. In the consent form that was completed prior to the study session, participants were told that the purpose of the study was to evaluate and improve the design of a new voice assistant technology that was ready for market. After providing their consent, participants enrolled in an online study session, conducted via the Zoom video conferencing platform by a member of the research team. Half of the participants were randomly assigned to a low error rate condition, and half of the participants were randomly sorted into a high error rate condition. This random assignment occurred before beginning the study procedure. As described below, all participants completed the same basic set of tasks in interacting with the voice assistant; however, based on their assigned experimental condition, the voice assistant's responses to participants' queries were pre-determined to exhibit either a higher or lower rate of errors of speech recognition on specific tasks in the sequence created for the study.

\begin{figure}
    \centering
    \subfloat[\centering Example slide 1: "Imagine you are planning lunch with your friend. Please use the assistant to find recommendations for restaurants in your city." Voice Assistant Response (Inaccurate): "I didn't understand what you said."]{{\includegraphics[width=6cm]{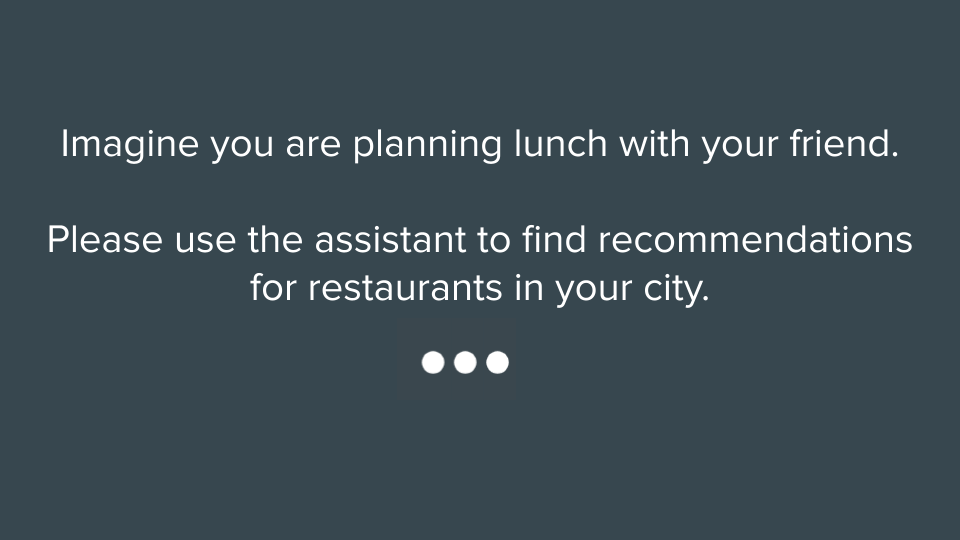} }}%
    \qquad
    \subfloat[\centering Example slide 2: "Please use the assistant to check the time." Voice Assistant Response (Accurate): "It is 00:00AM/PM" Response was adjusted manually by the researcher based on the participants' local time-zone.]{{\includegraphics[width=6cm]{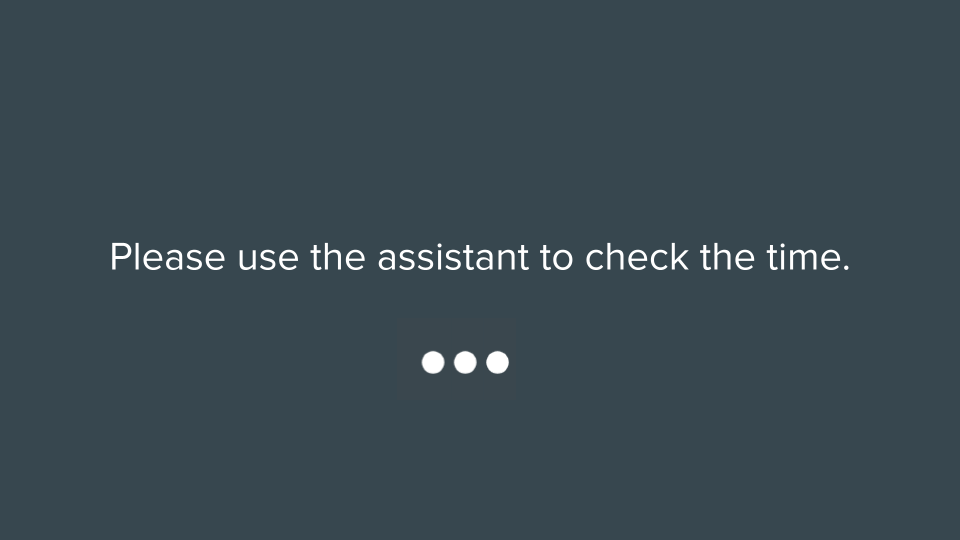} }}%
    \caption{Examples of slides screenshared with participants during the Wizard of Oz experimental procedure. The three dots on the slides would buffer animate once ``voice activated.''}%
    \label{fig:screenshots}
\end{figure}

After confirming participants' identity, compensation method, and consent, researchers turned off their web cameras and shared a slide show in full screen. Each slide featured different prompts instructing participants on how they should interact with the voice assistant (Figure \ref{fig:screenshots}). Participants were instructed to activate the voice assistant by saying ``Hey assistant'' before making any requests, and to speak to the assistant using a natural dialogue like they would use with their own voice assistant in their everyday life. Using a ``Wizard of Oz'' method, the researchers manually delivered all responses from the voice assistant using the text-to-speech AI voice generator from Play.ht \cite{playht}. The researchers aimed to replicate the default user experience of popular commercial products, and thus selected a voice representative of a woman speaking white American English \cite{moran2021racial}. Participants engaged with three ``warm-up'' prompts to get situated with the VA (Appendix Table \ref{tab:commands}) before users were presented with eleven evaluative prompts (Table \ref{tab:response}). The prompts were selected based on prominent VA user habits, as reported by a 2019 Adobe survey of over 1,000 users \cite{adobe} and system logs of voice assistant users' commands \cite{sciuto2018hey}. We implemented humorous VA responses in the beginning and end of each participant's VA interaction to make participants feel more comfortable and enhance their task enjoyment in the study environment \cite{niculescu2013making}. Prior to the study, this procedure was carefully piloted to ensure that participants perceived a high degree of realism and believed they were, in fact, interacting with a functioning voice assistant. 

Based on a participants' randomly assigned error rate condition, the audio response delivered by the  voice assistant would either accurately or inaccurately address the participants' requests. For participants in the high error rate condition, 35.7\% of the voice assistant responses were inaccurate. This error rate is based on prior research on the error rates Black individuals experience with voice assistants in everyday environments \cite{koenecke2020racial}. For participants in the low error rate condition, 7.1\% of the responses were inaccurate. We chose to implement an error rate lower than what white Americans typically experience as we were aiming to simulate an ideal version of the software. That said, we still included one inaccuracy in the low error rate condition, as no commercial voice assistant has perfect accuracy and we wanted our product to be accepted as a realistic product. To this point, in our pilot studies, participants who interacted with a voice assistant displaying perfect accuracy were more skeptical that the voice assistant was real, echoing previous research on agentic errors \cite{ragni2016errare, mirnig2017err}.

\begin{table*}
\caption{Transcription of the VA text prompts shared by researchers through a slide deck, and the responses the WoZ VA gave in the high and low word error rate (WER) conditions. For variable responses (i.e. regarding the weather, time, and Billboard charts), a sample response is included in the table. During the experiment, variable responses were appropriately changed by the researchers.}
\label{tab:response}
\resizebox{!}{.33\paperheight}{%
  \begin{tabular}{p{5.5cm}p{4.5cm}p{4.5cm}}
  \toprule
\textbf{On-Screen Text Prompt} & \textbf{VA High WER  Response} & \textbf{VA Low WER Response} \\
\midrule
\textbf{Imagine you have just started your day:} &  & \\
Please use the assistant to check the news. & \textit{{[}Reads 2 national headlines from that day{]}}  & \textit{{[}Reads 2 national headlines from that day{]}} \\
Please use the assistant to check the weather.  & Um, I didn't quite get that. & It's currently partly cloudy and 37 degrees in Chicago, Illinois, Expect snow starting tonight, today's high will be 39 degrees and the low will be 29. \\
Imagine you are planning lunch with your friend. Please use the assistant to find recommendations for restaurants in your city.  & I didn't understand what you said.  & I didn't understand what you said.  \\
Imagine you to tell a joke when you meet up with your friend. Please ask the assistant to tell you a joke. & What did the tree say to the moss?...(pause) You’re starting to grow on me. & What did the tree say to the moss?...(pause) You’re starting to grow on me.  \\
Imagine you’re getting ready for your meet-up, please use the assistant to play ‘Hello’ by Lionel Richie.  & Playing Hello by Adele. & Playing Hello by Lionel Richie.   \\
\midrule
\textbf{Imagine you are on your way to lunch:} & &  \\
Please use the assistant to check the time.  & It’s 5:17 PM.  & It’s 5:17 PM. \\
Please use the assistant to find out who won the Grammy for best album in 2021. & I don’t understand what you are saying  & The Grammy award for Best Album in 2021 went to Taylor Swift, for the album Folklore. \\
Please use the assistant to check the top songs on the Billboard charts.   & According to Billboard, the top songs on the Hot 100 today are Butter by BTS, Good For You by Olivia Rodrigo, and Levitating by Dua Lipa Featuring Da Baby & According to Billboard, the top songs on the Hot 100 today are Butter by BTS, Good For You by Olivia Rodrigo, and Levitating by Dua Lipa Featuring Da Baby \\
\midrule
\textbf{Imagine you are making pancakes from a recipe:} &   & \\
The recipe calls for 100 grams of flour, please use the assistant to convert 100 grams to ounces.   & There are 3.53 ounces in 100 grams.  & There are 3.53 ounces in 100 grams. \\
You just put your first pancake in the pan, please use the assistant to set a timer for 30 seconds to remind you to flip the pancake.  & 13 seconds starting now   & Setting a timer for 30 seconds.  \\
Please ask the assistant if they prefer blueberries or chocolate chips in her pancakes.   & I like microchip pancakes, I mean mint chocolate chip pancakes.    & I like microchip pancakes, I mean mint chocolate chip pancakes. \\
\end{tabular}%
}
\end{table*}

\subsection{Outcome Measures}
After completing their set of interactions with the voice assistant, participants completed a survey about their experience. The survey included the following validated self-report measures of their psychological responses to their experience as well as their perceptions of the technology: 

\subsubsection{Affective Responses} The PANAS-X \cite{watson1988development} was used to measure participants' state of positive and negative affect following their interaction with the voice assistant. This scale includes 60 individual items representing different positive emotions (e.g., cheerful, delighted, energetic) and negative emotions (e.g., irritable, upset, downhearted). Participants rated the extent to which they were experiencing each of these emotions using a 5-point Likert scale ranging from 1 (very slightly or not at all) to 5 (extremely).  

\subsubsection{Self-Consciousness} To measure participants' level of \textit{ self-consciousness} -- that is, their level of awareness of and focus on themselves -- we utilized a validated scale developed by Fenigstein and colleagues \cite{fenigstein1975public}. This scale is comprised of 9 statements, which participants utilize a 7-point Likert scale (anchored with the labels \textit{Strongly Disagree} and \textit{Strongly Agree}) to express their agreement that the statement accurately describes how they are currently feeling. Sample items include: ``Right now I am keenly aware of everything in my environment,'' ``Right now I am concerned about what other people think of me,'' and ``Right now, I am concerned about the way I present myself.''

\subsubsection{Self-Esteem} The Rosenberg Self-Esteem Scale \cite{rosenberg1965rosenberg, robins2001measuring} was used to measure individual state-level self-esteem. It contains 10 statements which participants rated using a 5-point Likert scale (\textit{Strongly Disagree} to \textit{Strongly Agree}). Sample items include: ``I take a positive attitude toward myself,'' ``I wish I could have more respect for myself,'' ``On the whole, I am satisfied with myself,'' and ``I feel I do not have much to be proud of.'' 

\begin{table*}[ht!]
\caption{Mean (M) and standard deviation (SD) for each participant condition and survey outcome measurement. Shading indicates a statistically significant difference in means between the low and high error rate (ER) groups for the respective race condition. Shaded rows for Black participants indicate \textit{p} < .01, and for white participants \textit{p} < 0.05.}
\begin{tabularx}{\textwidth}{ |c| *{8}{Y|} }
\cline{2-9}
   \multicolumn{1}{c|}{} 
 & \multicolumn{2}{c|}{\textbf{Black low ER}}  
 & \multicolumn{2}{c|}{\textbf{Black high ER}}
  & \multicolumn{2}{c|}{\textbf{white low ER}}  
 & \multicolumn{2}{c|}{\textbf{white high ER}} \\
\hline
 \textit{Dependent Variable} & \textbf{M} & \textbf{SD} & \textbf{M} & \textbf{SD} & \textbf{M} & \textbf{SD} & \textbf{M} & \textbf{SD}\\
\hline
\textit{PANAS-X Positive [1-5]}  & \cellcolor{shade}3.68 & 0.56 & \cellcolor{shade}2.74 & 0.94 & 3.20 & 0.86 & 3.18 & 0.92\\
 \hline
 \textit{PANAS-X Negative [1-5]}  & 1.33 & 0.31 & 1.75 & 0.75 & 1.19 & 0.29 & 1.45 & 0.39\\
 \hline
 \textit{Self-Consciousness [1-7]}  & \cellcolor{shade}4.94 & 0.75 & \cellcolor{shade}6.09 & 0.72 & 4.28 &  0.78 & 4.50 & 0.75\\
\hline
 \textit{Individual Self-Esteem [1-5]}  & \cellcolor{shade}4.99 & 0.59 & \cellcolor{shade}4.17 & 0.74 & \cellcolor{shade}4.90 & 0.68 & \cellcolor{shade}4.64 & 0.47\\
\hline
\textit{Collective Self-Esteem [1-5]}  & \cellcolor{shade}4.85 & 0.56 & \cellcolor{shade}4.26 & 0.68 & 4.45 & 0.56 & 4.57 & 0.59\\
\hline
\textit{Transportation [1-7]}  & 4.32 & 0.26 & 3.96 & 0.56 & 4.32 & 0.65 & 4.01 & 0.52\\
\hline
\textit{Tech Evaluation [1-7]}  & \cellcolor{shade}5.30 & 0.48 & \cellcolor{shade}4.46 & 1.14 & 4.77 & 0.94 & 5.06 & 0.68\\
\hline
\end{tabularx}
\label{tab:meanOutcome}
\end{table*}

To measure participants' perceptions of worth regarding their social identity, we employed the Collective Self-Esteem Scale \cite{luhtanen1992collective}. It contains 16 items measuring how people feel about their group membership (e.g., ``I am a worthy member of the social groups I belong to''), their private thoughts about their identity group (e.g., ``I often regret that I belong to some of the social groups I do''),  their  perceptions of external valuations of their identity group (e.g., ``In general, others respect the social groups that I am a member of''), and the importance of social identity groups to their sense of identity (e.g., ``The social groups I belong to are an important reflection of who I am'').

\subsubsection{Psychological Transportation} To measure participants' level of immersion and engagement with the VA during their interaction, we utilized an adapted version of the Transportation Scale \cite{green2000role}. This scale contains eight items assessing the degree of mental involvement in a specific task, with each item using a 7-point Likert scale (anchored with scale points labeled \textit{Strongly Disagree} and \textit{Strongly Agree}). Sample items include: "I was mentally involved in the experience" and "I found my mind wandering" (reverse-scored). 

\subsubsection{Evaluations of the Technology} To understand how participants felt about the VA that they interacted with during the experiment, we asked participants to rate the technology along eleven dimensions, each utilizing a 7-point semantic differentials scale anchored with opposing traits (e.g., useful-useless; beneficial-harmful; designed for me-not designed for me).   

\section{Results}

To analyze the results for each of the scales utilized in the post-interaction survey, we utilized a 2-factor analysis of variance (ANOVA), with participant race and the error rate condition as the independent variables. A Bonferroni correction was applied to control for family-wise type 1 error rate; all \textit{p}-values reported are adjusted for this correction. Based on our hypotheses, we expected to observe significant interactions between race and error condition on the outcome measures, which would indicate that the pattern of responses between the low and high error rates would differ between Black and white participants. Specifically, we predicted that Black participants would exhibit a more significant differentiation in response, in line with our prediction that Black, but not white, participants would experience stronger negative responses parallel to those demonstrated in prior research on racial microaggressions. Refer to Table \ref{tab:meanOutcome} for the mean outcome measures for all outcome variables and Figures 2-8 for data visualizations.

\begin{figure}
     \centering
     \captionsetup[subfigure]{labelformat=empty}
     \begin{subfigure}[b]{\columnwidth}
         \centering
         \includegraphics[width=\textwidth]{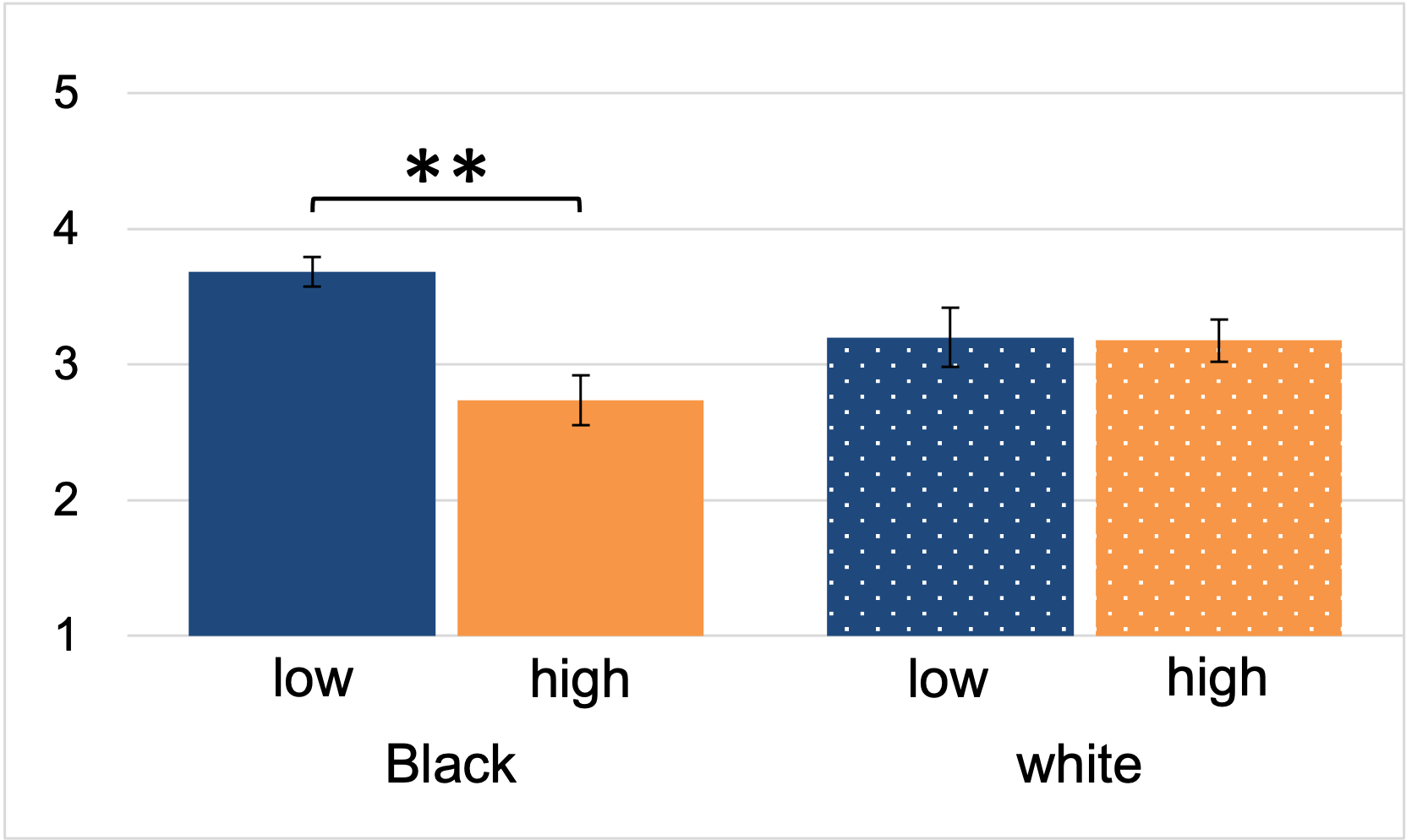}
         \caption{PANAS-X Positive Affect}
         \label{fig:panasPos}
     \end{subfigure}
     \caption{Barplot of PANAS-X Scale positive outcome means with standard error bars. Double asterisk (**) denotes a significant between-conditions difference (\textit{p} < .01).
     Black participants in the high error rate condition exhibited significantly lower levels of positive affect than Black participants in the low error rate condition. We did not observe this difference in white participants.}
     \Description[Barplot with 4 bars.]{5-point Y-axis representing the PANAS-X Positive Affect outcome measurement. X-axis representing the participant experimental conditions. The first two bars represent the low and high error rate conditions for Black participants, and are marked with a double asterisk. Black participants in the low error rate condition had a mean of 3.68. Black participants in the high error rate condition had a mean of 2.74. The second two bars represent the low and high error rate conditions for white participants. White participants in the low error rate condition had a mean of 3.20. White participants in the high error rate condition had a mean of 3.18.}
\end{figure}
\hfill
\begin{figure}
    \centering
    \captionsetup[subfigure]{labelformat=empty}
     \begin{subfigure}[b]{\columnwidth}
         \centering
         \includegraphics[width=\textwidth]{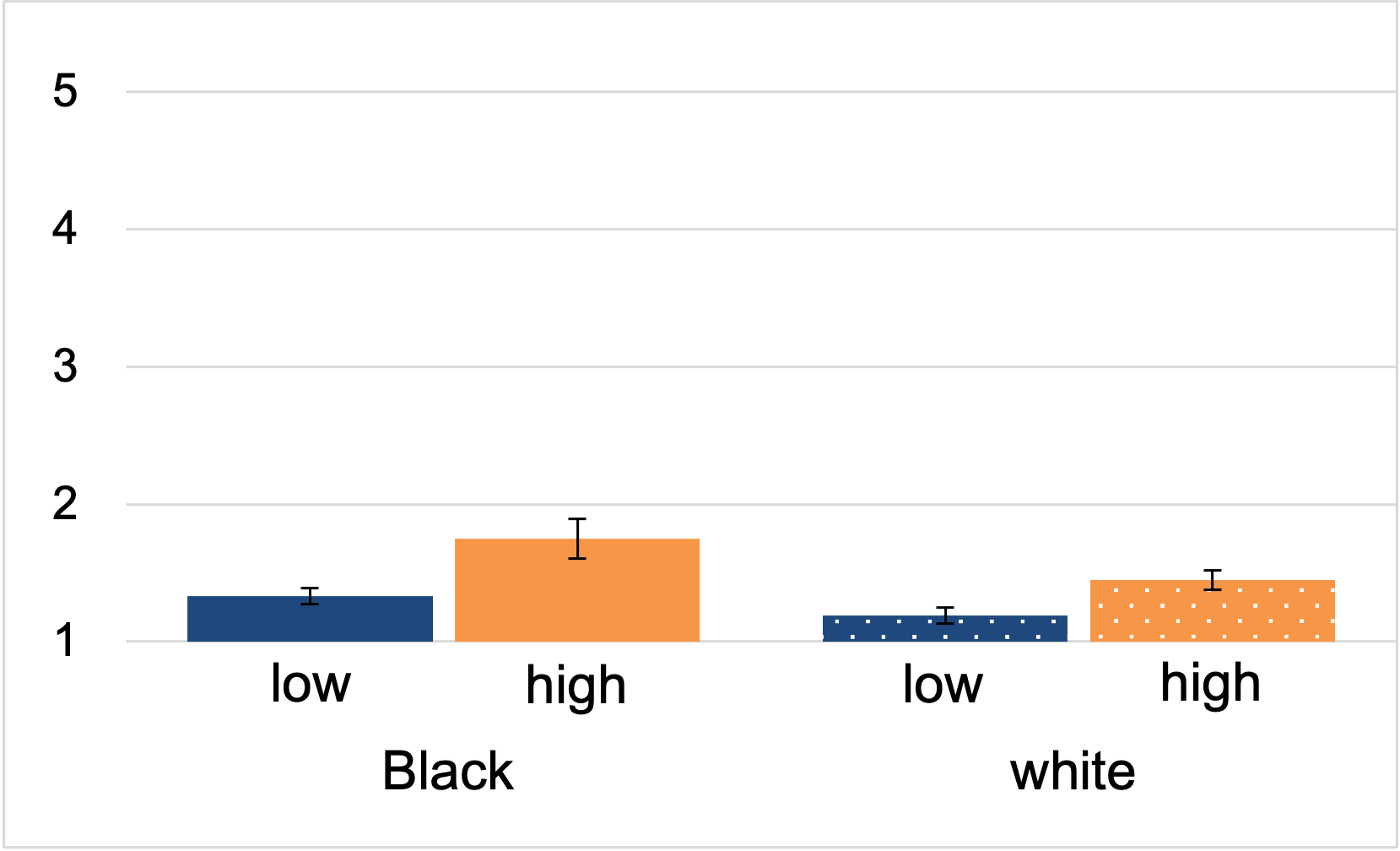}
         \caption{PANAS-X Negative Affect}
         \label{fig:panasNeg}
     \end{subfigure}
     \caption{Barplot of PANAS-X Scale negative outcome means with standard error bars. We did not observe a significant interaction effect in our negative affect measurement.}
     \Description[Barplot with 4 bars.]{5-point Y-axis representing the PANAS-X Negative Affect outcome measurement. X-axis representing the participant experimental conditions. The first two bars represent the low and high error rate conditions for Black participants. Black participants in the low error rate condition had a mean of 1.33. Black participants in the high error rate condition had a mean of 1.75. The second two bars represent the low and high error rate conditions for white participants. White participants in the low error rate condition had a mean of 1.19. White participants in the high error rate condition had a mean of 1.45.}
\end{figure}

\subsubsection{Affective Responses}
(Figure 2 and Figure 3) To analyze the results from the PANAS-X Scale of affective responses, we first created separate composite subscales for the Positive Affect and Negative Affect items; each subscale achieved a satisfactory level of internal reliability (Cronbach's alphas > 0.75). 

Results from the ANOVA for the Positive Affect scale revealed a significant race x error condition interaction: \textit{F} (1, 107) = 5.74, \textit{p} = .007. Planned comparisons revealed that Black participants in the high-error condition reported a significantly lower level of positive affect (\textit{M}= 2.74, \textit{SD} = .94) compared to Black participants in the low-error condition (\textit{M}= 3.68, \textit{SD} = .56), \textit{t}(52) = 4.47, \textit{p} < .01. In comparison, there was no significant difference in the average level of positive affect reported by white participants in the high-error condition (\textit{M}= 3.18, \textit{SD} = .92) and low-error condition (\textit{M}= 3.20, \textit{SD} = .86),  \textit{t}(52) = .82, \textit{p} = .47. This pattern supports our hypothesis that Black participants' positive emotional states would be more negatively affected by encountering a higher rate of errors than would white participants'. 

Analysis of the responses to the Negative Affect scale did not reveal a significant race x error condition interaction: \textit{F} (1, 107) = .17, \textit{p} = .39. Overall, reported levels of negative affect were relatively low (with means in all conditions falling below the midpoint of the 5-point rating scale). Average levels of negative affect were higher in the high-error conditions (\textit{M}= 1.59, \textit{SD} = .56) compared to the low-error conditions (\textit{M}= 1.26, \textit{SD} = .30), but this pattern did not differ by participant race. These results did not support our hypothesis: neither Black nor white participants appeared to experience a high level of negative affect overall.

\subsubsection{Self-Consciousness}
(Figure 4) Participants' responses to the individual items of the Self-Consciousness Scale were summed and averaged to form a composite score (Cronbach's alpha = .83). Results from the ANOVA for the composite scale revealed a significant race x error condition interaction: \textit{F} (1, 107) = 5.61, \textit{p} < .001. Planned comparisons revealed that Black participants in the high-error condition reported a significantly higher level of self-consciousness (\textit{M}= 6.09, \textit{SD} = .72) compared to Black participants in the low-error condition (\textit{M}= 4.94, \textit{SD} = .75), \textit{t}(52) = 7.42, \textit{p} < .01. In comparison, there was no significant difference in the average level of self-consciousness reported by white participants in the high-error condition (\textit{M}= 4.50, \textit{SD} = .75) and low-error condition (\textit{M}= 4.28, \textit{SD} = .78), \textit{t}(52) = .86, \textit{p} = .29. This pattern supports our hypothesis that Black participants' state of self-consciousness would be affected more by encountering a higher rate of errors than would white participants'.

\begin{figure}
    \centering
    \captionsetup[subfigure]{labelformat=empty}
    \begin{subfigure}[b]{\columnwidth}
         \centering
         \includegraphics[width=\textwidth]{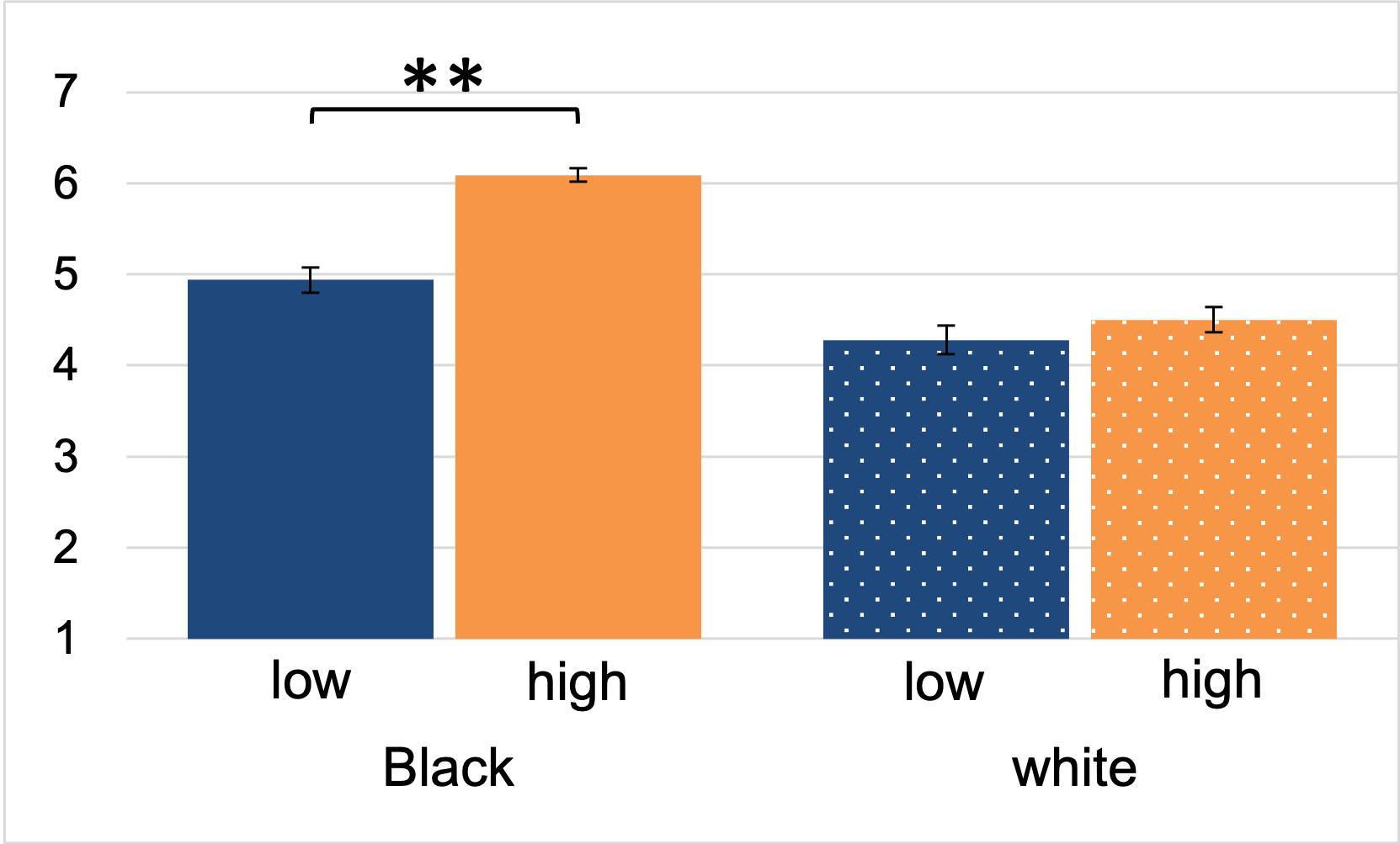}
         \caption{Self-Consciousness}
         \label{fig:selfconscious}
     \end{subfigure}\par\medskip
     \caption{Barplot of Self-Consciousness outcome means with standard error bars. Double asterisk (**) denotes a significant between-conditions difference (\textit{p} < .01). Black participants in the high error rate condition exhibited significantly higher levels of self-consciousness compared to Black participants in the low error rate condition. We did not observe this difference in white participants.} 
     \Description[Barplot with 4 bars.]{7-point Y-axis representing the Self-Consciousness outcome measurement. X-axis representing the participant experimental conditions. The first two bars represent the low and high error rate conditions for Black participants, and are marked with a double asterisk. Black participants in the low error rate condition had a mean of 4.94. Black participants in the high error rate condition had a mean of 6.09. The second two bars represent the low and high error rate conditions for white participants. White participants in the low error rate condition had a mean of 4.28. White participants in the high error rate condition had a mean of 4.50.}
\end{figure}

\begin{figure}
    \centering
    \captionsetup[subfigure]{labelformat=empty}
     \begin{subfigure}[b]{\columnwidth}
         \centering
         \includegraphics[width=\textwidth]{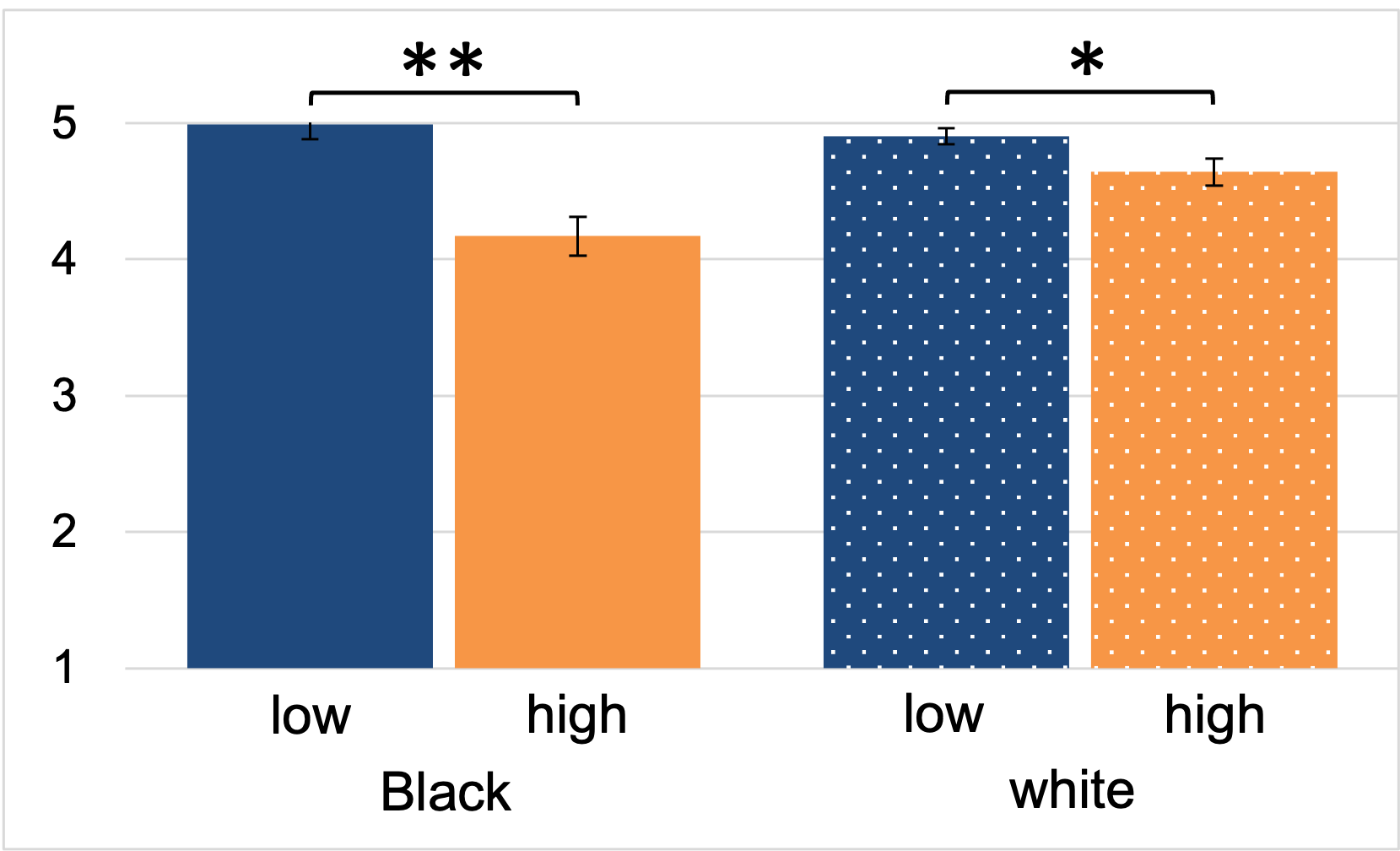}
         \caption{Individual Self-Esteem}
         \label{fig:individual}
     \end{subfigure}
     \caption{Barplot of Individual Self-Esteem outcome means with standard error bars. Double asterisk (**) denotes a significant between-conditions difference of (\textit{p} < .01), and single asterisk (*) denotes a significant between-conditions difference of (\textit{p} < 0.05). Black participants in the high error rate condition exhibited significantly lower individual self-esteem compared to Black participants in the low error rate condition. White participants also exhibited a significant difference in individual self-esteem across error rate conditions, although this difference was smaller than the difference we observed in Black participants.}
    \Description[Barplot with 4 bars.]{5-point Y-axis representing the Individual Self-Esteem outcome measurement. X-axis representing the participant experimental conditions. The first two bars represent the low and high error rate conditions for Black participants, and are marked with a double asterisk. Black participants in the low error rate condition had a mean of 4.99. Black participants in the high error rate condition had a mean of 4.17. The second two bars represent the low and high error rate conditions for white participants, and are marked with a single asterisk. White participants in the low error rate condition had a mean of 4.90. White participants in the high error rate condition had a mean of 4.64.}
 \end{figure}
 
 \begin{figure}
    \centering
    \captionsetup[subfigure]{labelformat=empty}
     \begin{subfigure}[b]{\columnwidth}
         \centering
         \includegraphics[width=\textwidth]{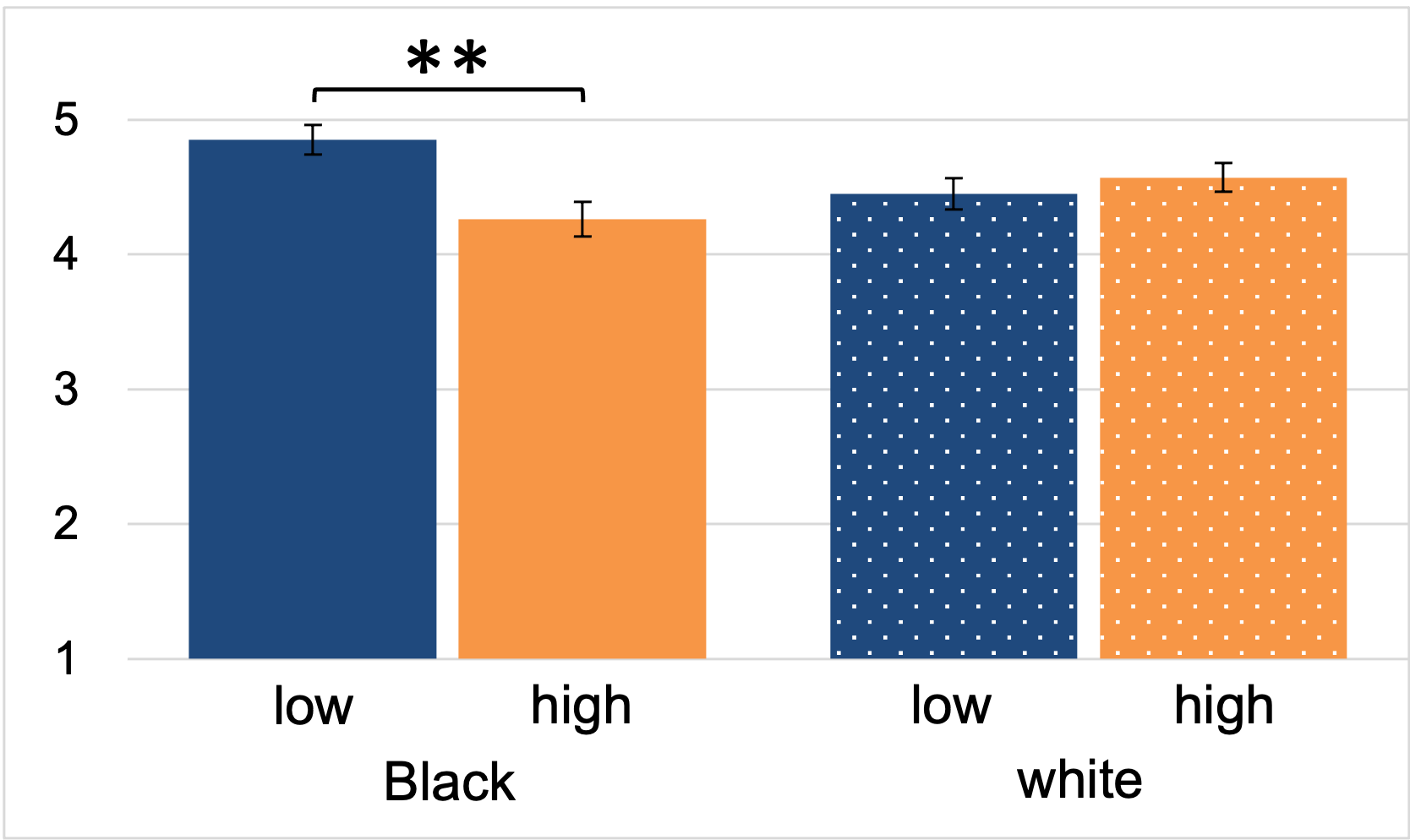}
         \caption{Collective Self-Esteem}
         \label{fig:collective}
     \end{subfigure}
     \caption{Barplot of Collective Self-Esteem outcome means with standard error bars. Double asterisk (**) denotes a significant between-conditions difference (\textit{p} < .01). Black participants in in the high error rate condition exhibited significantly lower collective self-esteem compared to Black participants in the low error rate condition. We did not observe this difference in white participants.}
     \Description[Barplot with 4 bars.]{5-point Y-axis representing the Collective Self-Esteem outcome measurement. X-axis representing the participant experimental conditions. The first two bars represent the low and high error rate conditions for Black participants, and are marked with a double asterisk. Black participants in the low error rate condition had a mean of 4.85. Black participants in the high error rate condition had a mean of 4.26. The second two bars represent the low and high error rate conditions for white participants. White participants in the low error rate condition had a mean of 4.45. White participants in the high error rate condition had a mean of 4.57.}
\end{figure}

\subsubsection{Self-Esteem}
 (Figure 5 and Figure 6) Responses to both the individual and collective Self-Esteem scales were averaged to form composite scores for each (Cronbach's alphas > .78). Results from the ANOVA for the composite scale for individual self-esteem revealed a significant race x error condition interaction: \textit{F} (1, 107) = 2.18, \textit{p} = .01. Planned comparisons revealed that Black participants in the high-error condition reported a significantly lower level of individual self-esteem (\textit{M}= 4.17, \textit{SD} = .74) compared to Black participants in the low-error condition (\textit{M}= 4.99, \textit{SD} = .59), \textit{t}(52) = 4.52, \textit{p} < .01. The average level of self-esteem reported by white participants in the high-error condition was also lower (\textit{M}= 4.64, \textit{SD} = .47) than the average level reported by white participants in the high error rate condition (\textit{M}= 4.90, \textit{SD} = .68), \textit{t}(52) = 2.06, \textit{p} =.04. However, the difference in means still stands to be greater for Black participants than for white participants, supporting our hypothesis that Black participants' personal self-esteem would be affected more by encountering a higher rate of errors than would white participants.

For collective self-esteem, results from the ANOVA revealed a significant race x error condition interaction: \textit{F} (1, 107) = 3.38, \textit{p} = .003. Planned comparisons revealed that Black participants in the high-error condition reported a significantly lower level of individual self-esteem (\textit{M}= 4.26, \textit{SD} = .68) compared to Black participants in the low-error condition (\textit{M}= 4.84, \textit{SD} = .56), \textit{t}(52) = 3.49, \textit{p} < .01. In comparison, there was no significant difference in the average level of collective self-esteem reported by white participants in the high-error condition (\textit{M}= 4.57, \textit{SD} = .59) and low-error condition (\textit{M}= 4.45, \textit{SD} = .56), \textit{t}(52) = .76, \textit{p} = .45. This pattern supports our hypothesis that Black participants' group-level self-esteem would be affected more by encountering a higher rate of errors than would white participants'. 

\begin{figure}
\centering
    \captionsetup[subfigure]{labelformat=empty}
     \begin{subfigure}[b]{\columnwidth}
         \centering
         \includegraphics[width=\textwidth]{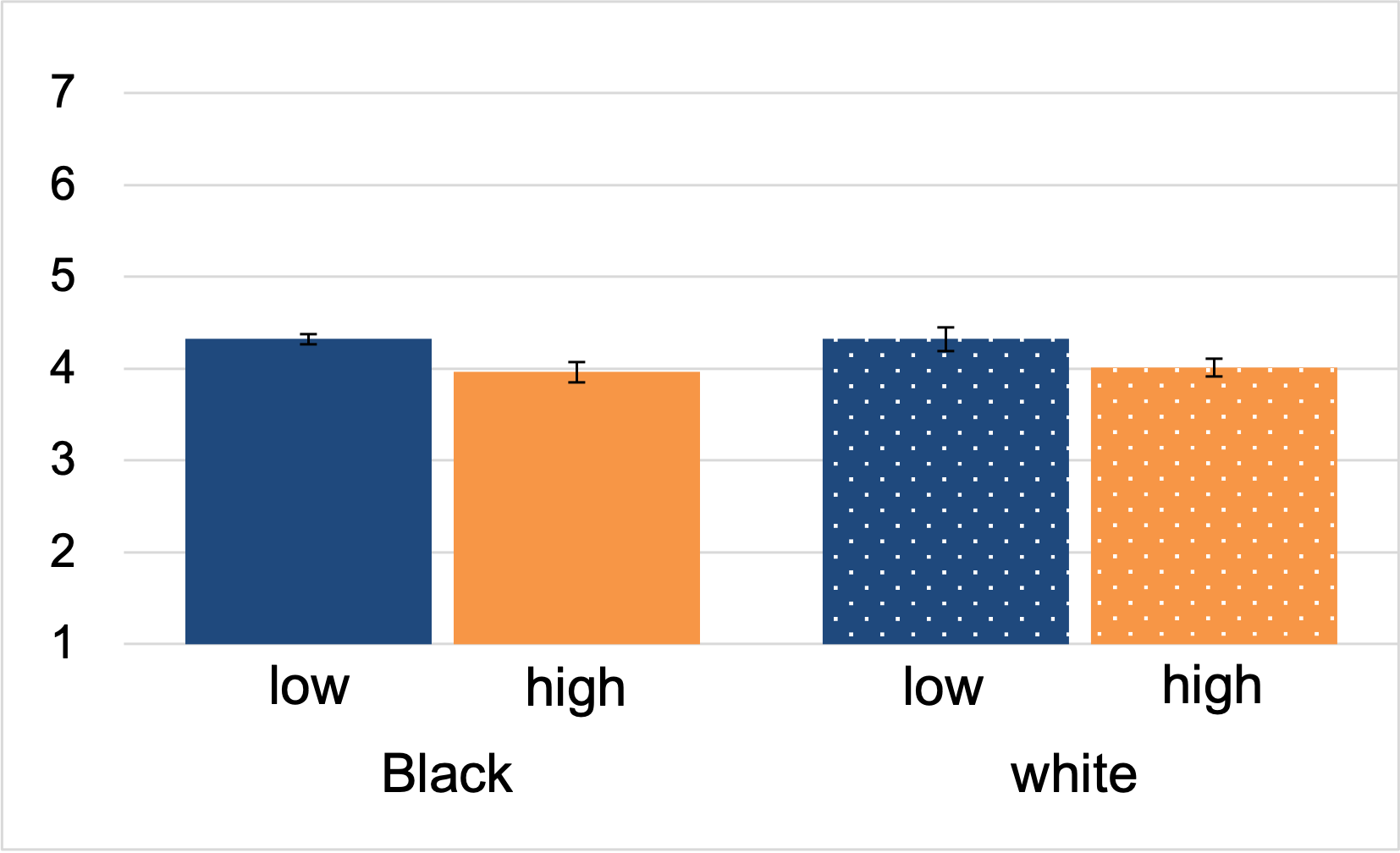}
         \caption{Transportation}
         \label{fig:transportation}
     \end{subfigure}
             \caption{Barplot of Transportation outcome means with standard error bars. No significant differences emerged between the high and low error conditions for either Black or white participants.}
            \Description[Barplot with 4 bars.]{7-point Y-axis representing the Transportation outcome measurement. X-axis representing the participant experimental conditions. The first two bars represent the low and high error rate conditions for Black participants. Black participants in the low error rate condition had a mean of 4.32. Black participants in the high error rate condition had a mean of 3.96. The second two bars represent the low and high error rate conditions for white participants. White participants in the low error rate condition had a mean of 4.32. White participants in the high error rate condition had a mean of 4.01.}
\end{figure}
\begin{figure}
         \centering
         \captionsetup[subfigure]{labelformat=empty}
     \begin{subfigure}[b]{\columnwidth}
         \includegraphics[width=\textwidth]{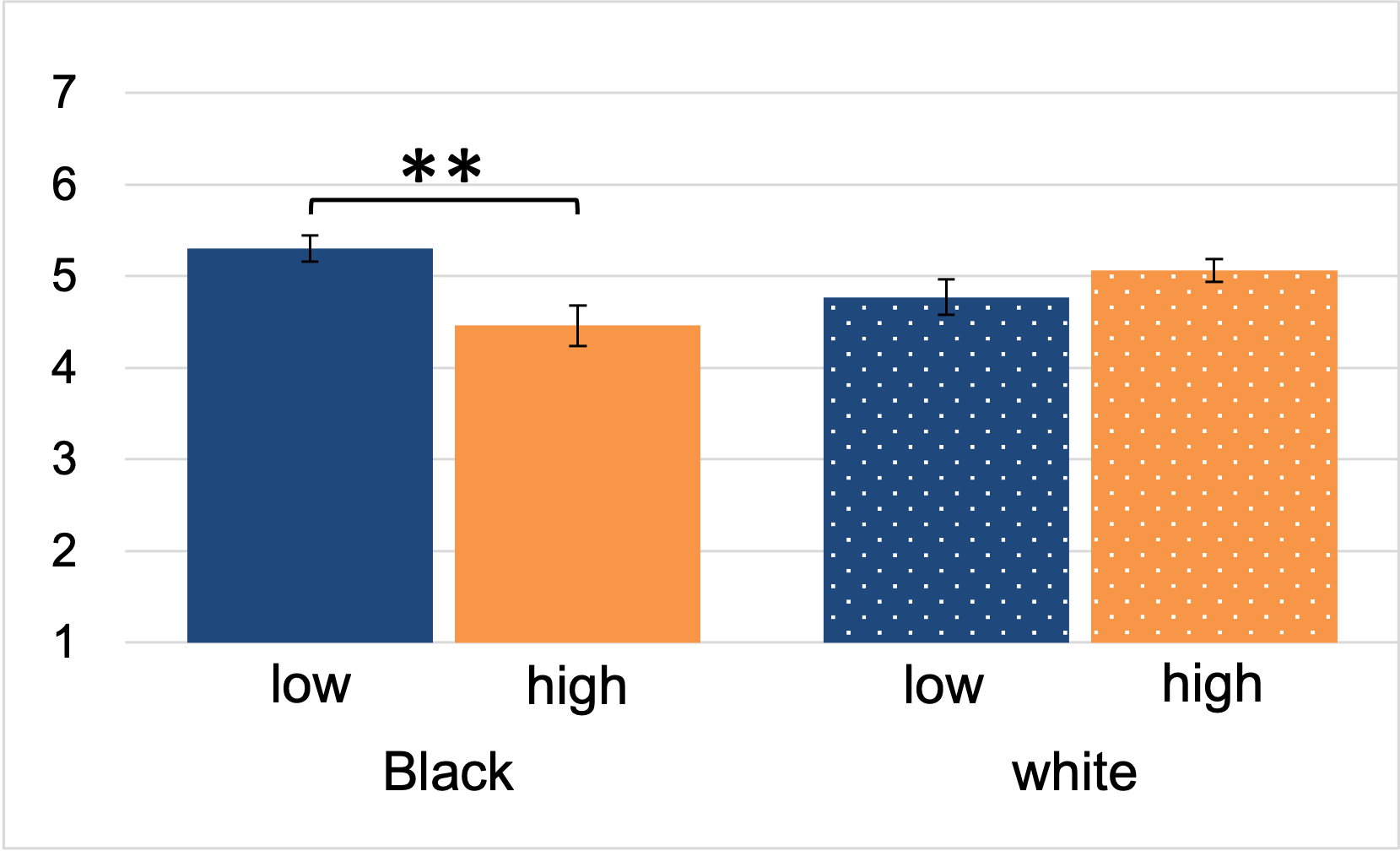}
         \caption{Tech Evaluation}
         \label{fig:techevaluation}
     \end{subfigure}
        \caption{Barplot of Tech Evaluation outcome means with standard error bars. Double asterisk (**) denotes a significant between-conditions difference (\textit{p} < .01). Black participants in the high error rate condition had significantly lower evaluations of the voice assistant compared to Black participants in the low error rate condition. We did not observe this difference in white participants.}
        \Description[Barplot with 4 bars.]{7-point Y-axis representing the Evaluations of Technology outcome measurement. X-axis representing the participant experimental conditions. The first two bars represent the low and high error rate conditions for Black participants, and are marked with a double asterisk. Black participants in the low error rate condition had a mean of 5.30. Black participants in the high error rate condition had a mean of 4.46. The second two bars represent the low and high error rate conditions for white participants. White participants in the low error rate condition had a mean of 4.77. White participants in the high error rate condition had a mean of 5.06.}
        \label{fig:transtech}
\end{figure}

\subsubsection{Transportation}
(Figure 7) Participants' responses to the individual items of the Transportation Scale were summed and averaged to form a composite score (Cronbach's alpha = .88). Results from the ANOVA for the composite scale revealed that the race x error condition interaction was not significant: \textit{F} (1, 107) = .01, \textit{p} = .82. Transportation levels reported by participants in the high-error conditions (\textit{M}= 3.99, \textit{SD} = .54) were lower than the mean levels reported in the low-error conditions (\textit{M}= 4.32, \textit{SD} = .48) to a non-significant degree, and this pattern held for both Black and white participants (see Table \ref{tab:meanOutcome}). Contrary to our hypotheses, Black participants did not show a differential rate of reduced engagement with the task, compared to white participants, when confronted with a more error-prone assistant. 

\subsubsection{Evaluations of the Technology}
(Figure 8) Participants' responses to the individual items of the technology evaluation measure were summed and averaged to form a composite score (Cronbach's alpha = .72). Results from the ANOVA for the composite scale revealed a significant race x error condition interaction: \textit{F} (1, 107) = 8.52, \textit{p} < .001. Planned comparisons revealed that Black participants in the high-error condition reported a significantly less positive evaluation of the voice assistant (\textit{M}= 4.46, \textit{SD} = 1.14) compared to Black participants in the low-error condition (\textit{M}= 5.30, \textit{SD} = .48), \textit{t}(52) = 3.51, \textit{p} < .01. In comparison, there was no significant difference in the average level of self-consciousness reported by white participants in the high-error condition (\textit{M}= 5.06, \textit{SD} = .68) and low-error condition (\textit{M}= 4.77, \textit{SD} = .94), \textit{t}(52) = 1.30, \textit{p} = .20. This pattern supports our hypothesis that Black participants' subjective perceptions of the technology would be more negatively impacted by interacting with a more error-prone VA than would white participants' perceptions. The pattern of means actually revealed that white participants rated the technology slightly (but not significantly) more \textit{positively} in the high-error condition.  

\section{Discussion}
\subsection{Summary of Results}
\label{summary}
Taken as a whole, the findings provide strong support for our general hypothesis that Black participants would be more negatively impacted by interacting with a more error-prone voice assistant than would white participants -- and, moreover, be impacted in ways consistent with findings from prior research on racial microaggressions. As the results of the study revealed, Black participants randomly assigned to the high-error condition, compared to Black participants in the low-error condition, exhibited higher levels of self-consciousness; lower levels of positive affect as well as individual and collective self-esteem; and less favorable evaluations of the technology. In contrast, white participants were largely unaffected by the error rate displayed by the assistant; across most measures, white participants displayed little difference in their psychological and evaluative responses. Moreover, the differences that were observed between Black and white participants, particularly in the high-error conditions, cannot be attributed to differences in engagement with the task (as we did not observe a significant race x error condition interaction for the measure of psychological transportation). 

In other words, despite the fact that white and Black participants in the high error condition experienced an objectively identical set of errors, their subjective experience of the interaction was strikingly different. This pattern is entirely consistent with the findings of prior work on racial microaggressions, which has revealed that the same life experiences (including being misunderstood or misinterpreted by others in social interactions) impact members of racial minority groups more negatively because those occurrences remind members of those groups of stereotypes or biases associated with their identity and trigger a host of threat-related emotional and cognitive responses. Linguistic and communicative misunderstandings are more systemic for Black individuals, but not for white individuals. Moreover, for many people of color, interpersonal microaggressions are constant, continual, and cumulative \cite{sue2019microaggressions}. The results from the present work indicate that people of color are likely to be affected similarly by acts of bias exhibited by technology and experience those interactions as microaggressions. Due to their innate racial privilege, white participants' race is not implicated in the same way in experiences of misunderstandings (by other people or by technology). Thus, instead of interpreting speech recognition errors as discriminating against their race or personhood, they are more likely to attribute the errors to other external factors \cite{torino2018everything}. Indeed, the pattern of Black participants' \textit{internalizing} the experience of VA errors (e.g., with heightened self-consciousness and reduced self-esteem) can be contrasted with the finding that white participants exhibited minimal patterns of self-directed focus or blame when confronted with the same display of misunderstanding from the VA. On the one dimension that white participants did appear to be negatively affected by VA errors, individual self-esteem, the impact was nonetheless significantly greater for Black participants.

\subsection{Limitations and Future Work}
The present study was designed to be an initial investigation of the disparate impact of voice assistant errors on marginalized and non-marginalized participants. The focus of the study was modeled on the prototype offered by controlled experimental research of racial microaggressions in its prioritization of a high level of experimental control and internal validity (e.g., in pre-designating interaction tasks and keeping the task sequences uniform between conditions), its focus on general differences between two demographic identity categories (Black versus white racial identity), and its use of validated outcome measures utilized by prior work in this space. At the same time, we acknowledge the limitations that these methodological choices pose and the value of follow-up work to extend the results the present study revealed.

First, in using a carefully controlled experimental set-up, we prioritized internal over external validity. While we were careful to design the VA interaction in ways that preserved a sense of believability and realism, this study did not deploy a manipulation check for realism and did not observe users' interactions with VAs in naturalistic settings. To this end, we have initiated a follow-up study utilizing in-the-wild data collection (including diary entries and usage logs) with participants in their own personal contexts to ascertain if the patterns of findings observed in the present research replicate in more natural, realistic interactions with VAs. 

Furthermore, this follow-up study aims to address a second limitation of the present work: its focus on the immediate, short-term psychological impact of VA errors on Black users. In the field study we are currently conducting, we are utilizing repeated measurement of many of the same outcome measures employed in the present study. In addition, we will incorporate a number of measures used in prior work on microaggressions to determine if repeated, cumulative experiences with biases in voice technologies affect users' susceptibility to health outcomes such as depression \cite{nadal2014impact, tynes2008online}, anxiety \cite{tynes2008online}, and an overall negative view of the world \cite{nadal2014impact}. Moreover, as researchers have demonstrated, repeated experiences with microaggressions and stereotype threat can have a host of physical health costs \cite{nadal2017injurious}, including high blood pressure \cite{blascovich2001african, brewer2013association} and hypertension \cite{roberts2008cross}. Future studies that utilize longitudinal studies should incorporate these longer-term measures of harm to determine the extent to which technology-driven microaggressions have a similar negative effect on people of color and other marginalized populations. 
In addition, future investigations, particularly longitudinal studies, could focus on the strategies use to respond to errors in technology -- for example, studying what factors predict particular behavioral responses to speech recognition errors, such as code-switching (i.e., assimilation to adjust speech to align with white American English: \cite{harrington2022s, kim2006reasons} or dis-engagement from interacting with error-prone technologies \cite{kuntsman2019paradox} and how such patterns of response might either exacerbate or mitigate any harm caused by a technology's performance.

Another inherent limitation of the present work is its focus on a single facet of identity -- racial identity -- and, moreover, its comparison of participants who identified their racial identity as primarily Black or white. Future work in this space must not only extend this finding to other facets of identity that may be susceptible to harm caused by patterns of bias in technology -- including other racial minority groups, other language groups (e.g., English as a second language speakers, speakers with particular accents or dialects), speakers from lower socio-economic strata, LGBTQ+ users, etc. Ideally, future work will also apply an intersectional approach to identity, understanding that the subjective experiences of individuals are impacted by the interplay between various facets of their identity \cite{rankin2019straighten}. For example, the mental and physical health implications of errors and biases in interactions with technology may be of particular significance for disabled Black users \cite{dunn2015person}. Since speech recognition technologies are utilized by individuals with a variety of accessibility needs \cite{shadiev2014review, wald2008universal, pradhan2018accessibility, azenkot2013exploring}, when these systems fail, not only are disabled Black users prevented from using assistive technologies that may be central to their day-to-day needs and workflow, but simply attempting to use these requisite technologies can increase their risk of suffering mental and physical health harms due to the psychological threat they may evoke.

Finally, the present research utilized a VA whose voice exhibited the typical features commonly used as the default in the most popular options on the market (e.g., Alexa, Siri, or Google Home): namely, a female voice that prior work has shown is assigned a racial identity of white \cite{moran2021racial}. Building on a growing body of work examining how various characteristics of voice assistants may affect user trust and acceptance, which has focused primarily on perceived gender \cite{goodman2023s, tolmeijer2021female, rincon2021speaking} and personality \cite{braun2019your, poushneh2021humanizing}, understanding the role of perceived \emph{race} of a VA would be a worthwhile focus for future work. For example, one specific follow-up study to the present research could manipulate both the error rate and perceived race of a VA to determine how users respond to an error-prone VA who shared versus does not share their own racial identity. While prior work has shown that Black users exhibited a preference for conversational agents perceived to be Black \cite{liao2020racial}, would perceived race impact the extent to which Black users experience a VA's speech recognition errors as a microaggression? 

\subsection{Designing for Harm Mitigation and Reduction}
Given the findings of the present study, one vital implication for the design of voice assistants is the importance of addressing or reversing any harm caused by errors in speech recognition, particularly for users from marginalized groups. While there is a growing body of work dedicated to understanding VA error recovery \cite{jiang2013users, myers2018patterns, motta2021users, mavrina2022alexa, cho2020role, beneteau2019communication}, little attention has been paid to how error recovery may be designed specifically for members of marginalized populations, such as Black users. Next, we propose potential directions for designing error recovery strategies that acknowledge the validity of marginalized users' experiences of speech recognition errors as microaggressions and/or aim to reduce the negative impact caused by these errors. These directions are directly informed by research on effective ways of defusing or mitigating the harm caused by experiences of bias or prejudice in everyday life \cite{sue2020microintervention}. 

\subsubsection{Coping with Microaggressions} 
\paragraph{Spot Checks} Oftentimes, people who have experiences that they perceive to be microaggressions are told that they are being ``too sensitive'' or that ``race has nothing to do with it'' \cite{sue2010microaggressions}. These messages are not only incorrect, as scholars have demonstrated time and time again that race is a prominent feature of linguistic discrimination \cite{haque2015indigenous, davila2016inevitability, flores2015undoing, rosa2019looking, rosa2017unsettling}, but they also diminish targets' experiences. Spot checks can help validate targets' experiential reality, and one way this can be achieved is to have a microaggressive act clearly identified and addressed in the context in which it occurs \cite{sue2010microaggressions, sue2020microintervention}. Some research has begun to explore how social technologies for people of color may involve elements of a spot check \cite{to2020they}; however the work to date has largely been speculative and, to our knowledge, no examples yet exist of a technology directly acknowledging its own inherent biases. In the context of an interaction with a voice assistant, this could take the form of the assistant acknowledging its limitations in accurately understanding the speech inputs from different identity groups and, equally important, validating the potential frustration and disappointment that users might feel if they are not well-understood.

\paragraph{Shifting Accountability} A related tactic that has been shown to be useful when responding to microaggressions people of color have experienced is ensuring that they do not place the blame of the act unto themselves. Acknowledging that the fault and responsibility of the microaggression lies in the perpetrator can help empower targets of acts of bias or discrimination \cite{sue2010microaggressions}. There has been some research on how virtual assistants may assume blame and and repair a conversation when an error occurs. For example, Cuadra \textit{et al.} found that when a VA makes a mistake, acknowledges its ownership of the mistake, and aims to repair the interaction (e.g., replies ``Hmm...It seems like I made a mistake, what's up?''), users respond more positively than when the VA acknowledges but does not take full ownership of the mistake (e.g., replies ``Sorry, I didn't get that''). Although around 20\% of that study's participants spoke English as a second language, the researchers did not focus on race as a factor in reporting or interpreting their results \cite{cuadra2021my}. How might a VA to reveal to users, following speech recognition errors, that its functionality is impacted by factors such as a lack of racial diversity in the voice data used to power its speech recognition capabilities? What form of acknowledgment and response would users from marginalized groups seek or desire in those instances? 

\paragraph{Identity Affirmations \& Collective Joy}
Other research has shown that affirming a positive aspect of one’s identity can counteract the negative effects of stereotype threat \cite{rydell2009multiple, martens2006combating}, and microaffirmations are beginning to emerge in clinical work to help patients combat microaggressions \cite{anzani2019absence, huber2021racial}. Affirmations provide a buffer to the psyche in the face of threat and can effectively reduce the harm to an individual's emotions or self-esteem following an ego-threatening experience -- for instance, by replacing thoughts related to stereotypes with thoughts that validate the worth and joy of one's identity \cite{logel2009perils}. Leveraging this line of research in the design of VA assistants could involve the technology following up a detected speech recognition error with an affirming question or message to the user. Based on this prior work, specific recommendations may include having a voice assistant include, in its acknowledgment of or follow-up to a speech recognition error, an expression of their general esteem for a user or an acknowledgment that the user relies on the assistant for information and aid with tasks and outcomes that are important to a user's everyday life. Such affirmations, while seemingly small, have been shown to provide a buffer to the threats to the ego posed by microaggressions.

\subsubsection{Designing with Marginalized Users}

The design directions we have proposed here are intentional in their focus on an \emph{assets-based} perspective on the experience of microaggressions and stereotype threat -- a perspective that recognizes marginalized individuals' unique cultural wealth and personal value \cite{hess2007educating, villalpando2005role}. An assets-based approach can be directly contrasted to a deficit-based approach, which casts members of marginalized groups as powerless or deficient, as it emphasizes that experiences that negatively impact members of marginalized groups are more a testament to the power of societal and situational forces that impact well-being \cite{morgan2007revitalising}. By emphasizing the importance of externalizing focus toward the perpetrating entity, and leveraging resources such as self-affirmation and joy, the design implications offered here aim to draw on the inner strength and resilience of members of marginalized groups. Moreover, we deliberately did not propose specific design ``solutions,'' as any reformulation of VA interactions should occur through participatory methods that engage and center the perspectives of marginalized groups \cite{harrington2019deconstructing}. 

\section{Conclusion}
Prior work in psychology has demonstrated the harmful psychological effects microaggressions and stereotype threat can have on people of color, and other research in HCI has documented the presence of bias in voice assistants. In this study, we synthesized these two phenomena to empirically study the psychological harm that bias in voice assistants may inflict on Black users. In addition to providing the first controlled experimental investigation of these effects, we aimed to inspire a host of future research through the research and design directions proposed.

\begin{acks}
We would like to thank Nik Martelaro for providing technical guidance in deploying our voice assistant. We would also like to thank Pranav Khadpe for reviewing pieces of this work and the research assistants who contributed to data collection, Kara Tippins and Yuchuan Shan. This work was supported by the National Science Foundation under Grant \#2040926.
\end{acks}

\bibliographystyle{ACM-Reference-Format}
\bibliography{references.bib}

\section{Appendix}
\begin{appendix}
\appendix
\section{Wizard of Oz Experiment, Researcher Script}
\label{appendix:script}
``During this study, you will be providing you with 11 questions you will be asking the voice assistant. The questions will be provided on screen as you advance through the study.
Due to this being over Zoom there might be a slight lag in the response time or feedback from the agent. This is completely normal. We have tested this and the voice agent accurately hears everything you say to it. 
Do you have any questions?

\emph{[Pause for any questions]}

As with any assistant, you must call on the assistant before asking it a question. For example with Google Home and Amazon Alexa, you would say ``Hey Google, or Alexa'' and then follow up with questions such as “What’s my schedule today?”. With this new voice technology, you'll also have to call on the assistant before asking it a question. To call on the assistant you can say ``Hey assistant'' and follow up with your question. 
If the voice assistant doesn't respond accurately or doesn't understand what you've asked, please refrain from re-asking the assistant. Furthermore, please answer all questions as naturally as possible, as if you were at home or in an environment where you regularly use your voice assistant. 
Lastly, after the study begins, I will remain in the background with my camera and mic off to encourage a seamless interaction between you and the assistant. Please refrain from asking me about any interactions between you and the assistant. 

Let’s run through a few questions to familiarize yourself with the assistant: \textit{[Refer to Table 3.]}

\begin{table*}
\caption{Transcription of the initial four VA text prompts shared by researchers through a slidedeck, and the responses the WoZ VA gave.}
\label{tab:commands}
  \begin{tabular}{p{5cm}p{5cm}}
  \toprule
\textbf{On-Screen Text Prompt} & \textbf{VA Response} \\
\midrule
\textbf{Please ask the assistant one of the following questions:} &  \\
Do you have any pets? & I don’t have any pets, I used to have a few bugs but they kept getting squashed. \\
What's your favorite sport?  & I’m more of a mathlete than an athlete.  \\
Do aliens exist? & So far there has been no proof that alien life exists but the universe is a very big place.  \\
What's your favorite color? & Yellow.  \\
\midrule
\textbf{Please ask the assistant to check how many feet are in a mile.} & There are 5,280 feet in a mile. \\
\midrule
\textbf{Please use the assistant to set an alarm for tomorrow at 3 pm.} & Your alarm is set for 3 PM tomorrow. \\
\midrule
\end{tabular}
\end{table*}

Do you have any questions or concerns?

\textit{[Pause for any questions]}

Great! You are now going to run through the bulk of the questions. 
Please imagine these conversations in the context of conversing with an agent at home or in an environment that you regularly use your voice assistant. 
I will now be turning off my camera and mic so you can converse with the assistant. I’ll pop back in after the questions are over.''

\end{appendix}

\end{document}